\documentclass[letter,traditabstract]{aa} 
\usepackage{graphicx}
\usepackage{txfonts}
\begin{document}
\title{{\it Herschel} Space Observatory\thanks{{\it Herschel} is an ESA space observatory with science instruments provided by European-led Principal Investigator consortia and with important participation from NASA.}}
\subtitle{An ESA facility for far-infrared and submillimetre astronomy}
\author{
	G.L. Pilbratt  \inst{1}\fnmsep\thanks{{\it Herschel} Project Scientist}
	\and
	J.R. Riedinger\inst{2}\fnmsep\thanks{{\it Herschel} Mission Manager}
	\and
	T. Passvogel\inst{3}
	\and
	G. Crone\inst{3}
	\and
	D. Doyle\inst{3}
	\and
	U. Gageur\inst{3}
	\and
	A.M. Heras\inst{1}
	\and \\
	C. Jewell\inst{3}
	\and 
	L. Metcalfe\inst{4}
	\and
	S. Ott\inst{2}
	\and
	M. Schmidt\inst{5}
}
\institute{ESA Research and Scientific Support Department, ESTEC/SRE-SA, Keplerlaan 1, NL-2201 AZ Noordwijk, 
	The Netherlands  \\  \email{gpilbratt@rssd.esa.int}
	\and
	    ESA Science Operations Department, ESTEC/SRE-OA, Keplerlaan 1, NL-2201 AZ Noordwijk, The Netherlands
         \and
             ESA Scientific Projects Department, ESTEC/SRE-P, Keplerlaan 1, NL-2201 AZ Noordwijk, The Netherlands
           	\and
	    ESA Science Operations Department, ESAC/SRE-OA, P.O. Box 78, E-28691 Villanueva de la Ca\~nada, Madrid, Spain
	\and
	    ESA Mission Operations Department, ESOC/OPS-OAH, Robert-Bosch-Strasse 5, D-64293 Darmstadt, Germany
}
\date{Received 9 April 2010; accepted 10 May 2010}
 \abstract{{\it Herschel} was launched on 14 May 2009, and is now an operational ESA space observatory offering unprecedented observational capabilities in the far-infrared and submillimetre spectral range 55$-$671\,$\mu$m. {\it Herschel} carries a 3.5 metre diameter passively cooled Cassegrain telescope, which is the largest of its kind and utilises a novel silicon carbide technology. The science payload comprises three instruments: two direct detection cameras/medium resolution spectrometers, PACS and SPIRE, and a very high-resolution heterodyne spectrometer, HIFI, whose focal plane units are housed inside a superfluid helium cryostat. 
{\it Herschel} is an observatory facility operated in partnership among ESA, the instrument consortia, and NASA. The mission lifetime is determined by the cryostat hold time. Nominally approximately 20,000 hours will be available for astronomy, 32\% is guaranteed time and the remainder is open to the worldwide general astronomical community through a standard competitive proposal procedure. 
 }
\maketitle

\section{Introduction}
\label{sec:intro}

The {\it Herschel} Space Observatory was successfully launched on 14 May 2009. However, it was conceived almost 30 years earlier as the Far InfraRed and Submillimetre Space Telescope (FIRST), which was formally proposed to ESA in November 1982 in response to a call for mission proposals issued in July 1982. After a feasibility study conducted in 1982-83, it was incorporated in the ESA `Horizon 2000' long-term plan (Longdon \cite{longdon1984}) for implementation as the third or fourth `cornerstone' mission. Over the years it was the subject of several studies adopting a variety of  mission designs, spacecraft configurations, telescopes, and science payloads. In November 1993 the ESA Science Programme Committee (SPC) decided that FIRST would be implemented as the fourth `cornerstone' mission. 

\begin{figure*}
    \centering{
	\includegraphics[height=82mm]{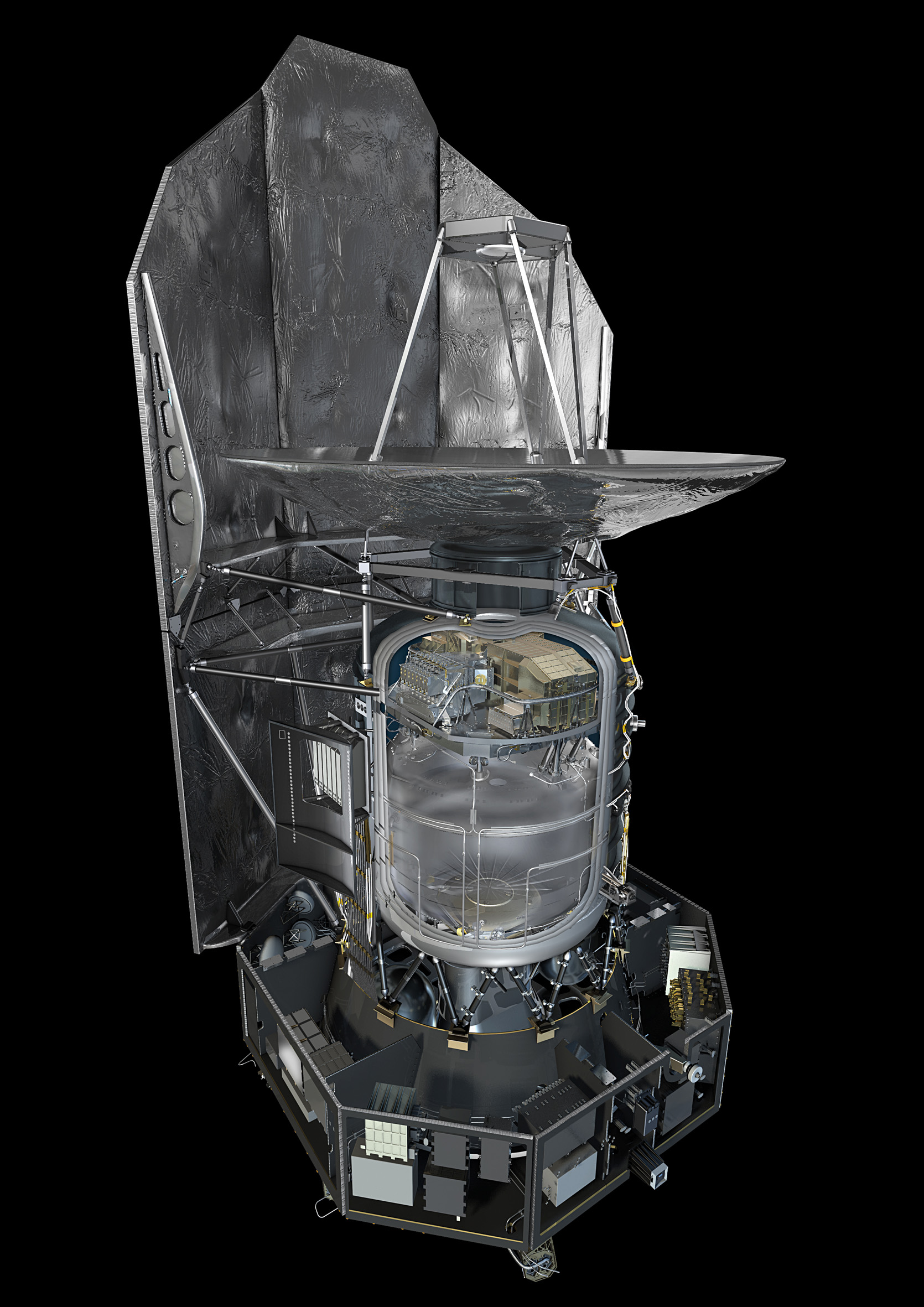}
	\includegraphics[height=82mm]{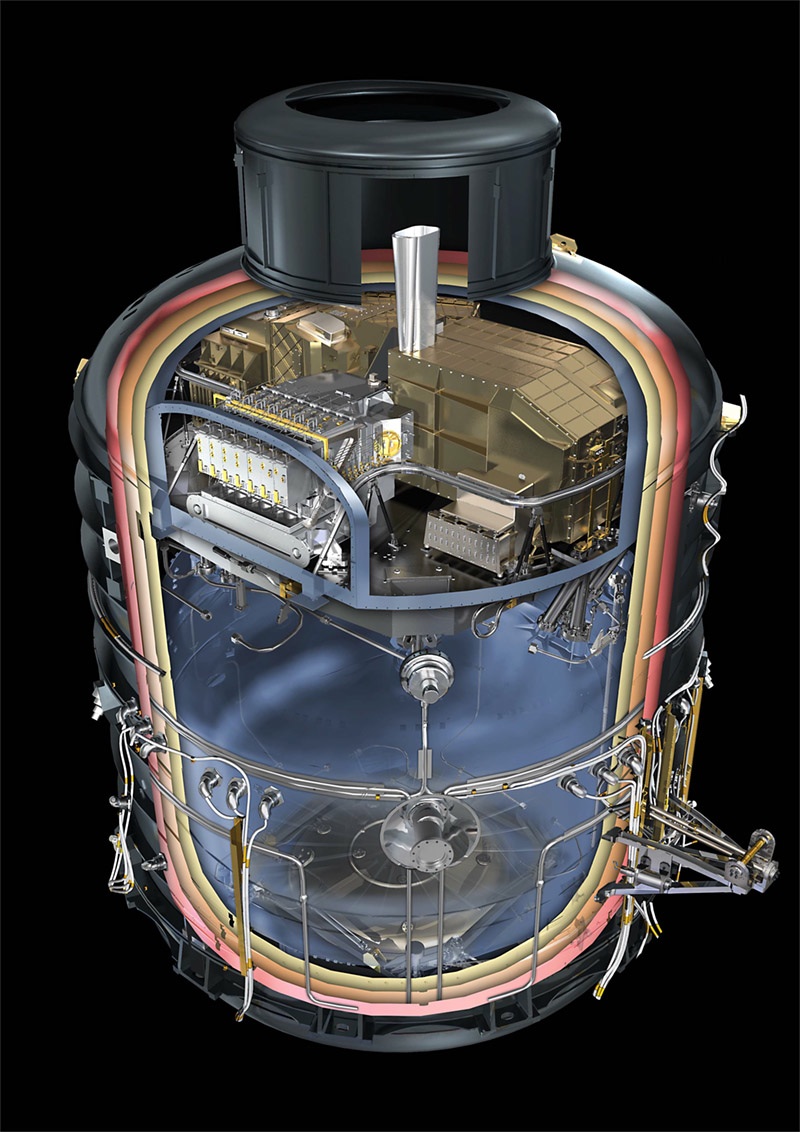}
	\includegraphics[height=82mm]{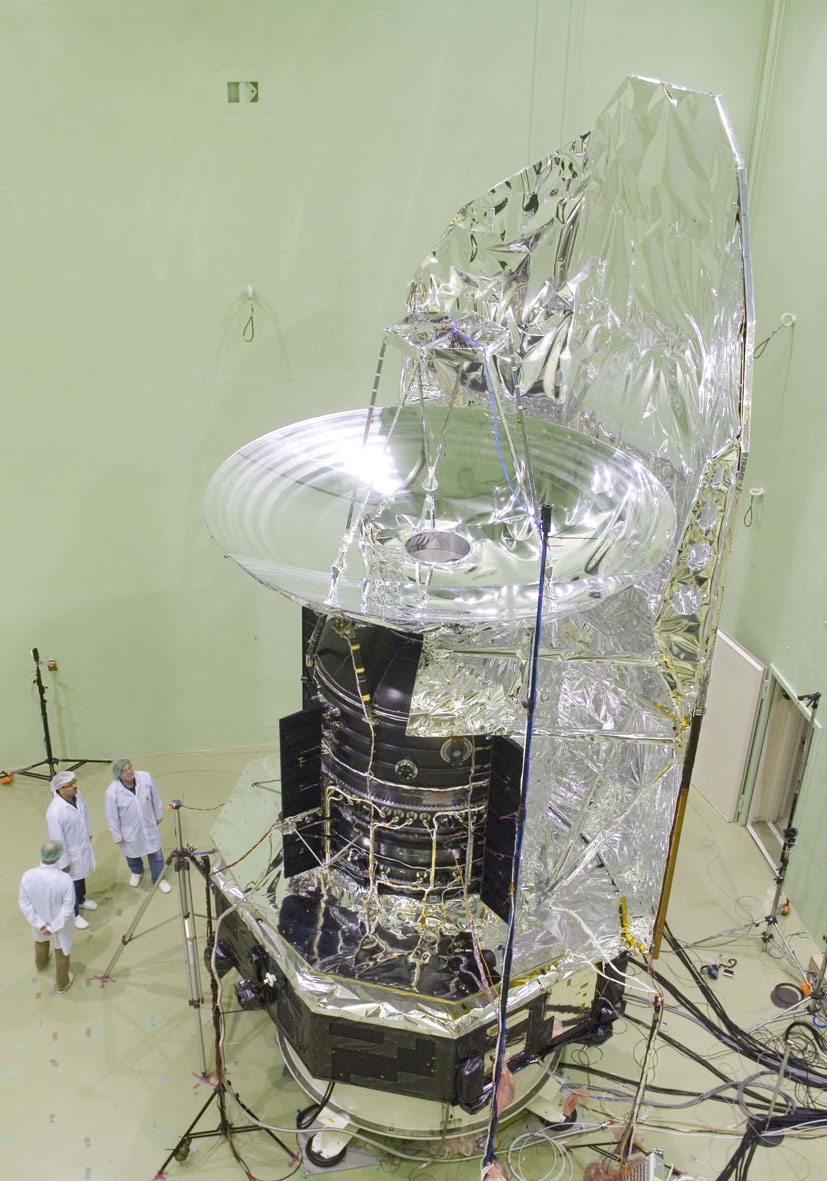}
  	}
    \caption{\label{fig:sc}
Left: The {\it Herschel} spacecraft features the `payload module' (PLM) with the cryostat housing the instrument focal plane units (FPUs), the telescope, the `service module' (SVM) with `warm' electronics, and the sunshield/sunshade. Middle: Close-up of the PLM displaying the optical bench with the instrument FPUs on top of the main helium tank. The focal plane cover and the three vapour-cooled shields inside the cryostat vacuum vessel (CVV) are also shown. Right: {\it Herschel} being prepared for acoustic testing in the Large European Acoustic Facility (LEAF) in the ESTEC Test Centre in June 2008, providing a good view of the telescope (ESA).
    }
\end{figure*}

The mission approved in 1993 was based on a spacecraft design with a 3\,m telescope passively cooled to 160\,K, two science instruments cooled by  mechanical coolers, operated in a near 24-hr period highly eccentric Earth orbit with continuous ground station coverage, and limited to conducting observations when outside the radiation belts. Through a critical reassessment of all aspects of the mission in the light of the experience gained from the Infrared Space Observatory (ISO, Kessler et al. \cite{kessler1996}) mission, the current mission concept with a spacecraft (Fig.\,\ref{fig:sc}) employing a superfluid helium cryostat reusing ISO technology, autonomously operating far away from the Earth in a large amplitude quasi-halo orbit around the 2nd Lagrangian point (L2) in the Sun-Earth/Moon system, eventually emerged. 

In October 1997 the Announcement of Opportunity (AO) for the science payload was released. Three instruments were provisionally selected in May 1998, and confirmed and approved by the SPC in February 1999. In September 2000 the Invitation to Tender (ITT) to industry for the spacecraft was issued, and in April 2001 the industrial activities commenced. Meanwhile, in December 2000 FIRST was renamed {\it Herschel} in recognition of the 200th anniversary of the discovery of infrared light by William Herschel (\cite{herschel1800}). 

The mission builds on earlier infrared space missions with cryogenic telescopes (the InfraRed Astronomical Satellite (IRAS, Neugebauer et al. \cite{neugebauer1984}), ISO,  {\it Spitzer} Space Telescope (Werner et al. \cite{werner2004}), and AKARI (Murakami et al. \cite{murakami2007})) and is designed to offer a larger telescope and to extend the spectral coverage further into the far-infrared and submillimetre, bridging the remaining relatively poorly explored spectral range to the radio astronomy space  (Submillimeter Wave Astronomy Satellite (SWAS, Melnick et al. \cite{melnick2000}) and Odin (Nordh et al. \cite{nordh2003}, Frisk et al. \cite{frisk2003})) and ground facilities.

The science rationale was extensively discussed in the community in a series of workshops, in particular in connection with the AO for the instruments (Rowan-Robinson et al. \cite{mrr1997}) and at the time of the ITT to industry (Pilbratt et al. \cite{glp2001}). 
About half of the energy emitted (typically at ultra-violet/optical/near-infrared wavelengths) in the universe since the epoch of recombination has been absorbed mainly by dust in the interstellar medium (ISM) in our Milky Way and other galaxies, and reradiated at much longer wavelengths. Stars in the early phases of formation in molecular clouds in the Galaxy and star-forming galaxies at redshifts up to z$\sim$5, covering the epochs of the bulk of the star formation in the universe, emit most of their energy in the {\it Herschel} spectral range. 

The prime science objectives of {\it Herschel} are intimately connected to the physics of and processes in the ISM in the widest sense. Near and far in both space and time, stretching from solar system objects and the relics of the formation of the sun and our solar system, over star formation in and feedback by evolved stars to the ISM,  to the star-formation history of the universe, galaxy evolution and cosmology.

\section{Mission design}
\label{sec:mission}

The {\it Herschel} mission adopted for implementation had the top level requirement to provide three years of routine science observations, employing a science payload to be provided by Principal Investigator (PI) consortia in exchange for guaranteed time (GT) observations.  
{\it Herschel} was to be operated as an observatory facility, comprising a  space segment - the satellite - and a ground segment, providing mission and science operations. 

\subsection{Spacecraft description}
\label{sec:sc}

The {\it Herschel} spacecraft (Fig.\ref{fig:sc} and Table\,\ref{tab:sc}) provides the appropriate working environment for the science instruments, points the telescope with required accuracy, autonomously executes the observing timeline, and performs onboard data handling and communication with the ground. It has a modular design, consisting of the `payload module' (PLM) supporting the telescope, the sunshade/sunshield, and the `service module' (SVM). The mission lifetime is determined by the cryostat lifetime, required to be 3.5 years, the initial six months were nominally allocated to early mission phases. 

\begin{table}[h]
\caption[]{Spacecraft and telescope main characteristics.}
\label{tab:sc} 
\centering
	 \begin{tabular}{ l  r }
	 \hline
	Spacecraft &   \\
	\hline
	Size height/width & ~~~~7.4 / 4.0 m \\
	Mass wet (incl. helium) / dry & 3400 (335) / 2800 kg \\
	(incl.  telescope/science instruments & 315 / 426 kg) \\ 
	Power total/science instruments & 1200 / 506 W \\
	Science data rate (max. average) & 130 kbps \\
	Solar aspect angle (wrt tel. boresight) & 60$\degr-$110$\degr$(120$\degr$) \\
	Absolute pointing (68\%) &  $\sim$2\arcsec \\
	\hline 
	Telescope &  \\
	\hline
	Primary physical/effective diameter & 3.5 / 3.28 m \\
	Secondary diameter	 &	30.8 cm  \\
	System/primary f-number & 8.70 / 0.5 \\
	Wave front error best-focus (centre/edge) & 4.8 / 5.5 $\mu$m \\
	Angular resolution &  $\sim$7$\arcsec\times(\lambda_\mathrm{obs}$/100 $\mu$m) \\
	Operational temperature & $\sim$85 K \\
	\hline
	\end{tabular}
\end{table}

The PLM is dominated by the cryostat vacuum vessel (CVV) from which the superfluid helium tank is suspended, surrounded by three vapour-cooled shields to minimise parasitic heat loads. The optical bench with the three instrument focal plane units (FPUs) is supported on top of the tank, which has a nominal capacity of 2367 litres. A phase separator allows a continuous evaporation of the liquid into cold gas. The FPUs and their detectors are kept at their required temperatures by thermal connections to the liquid cryogen in the tank and to pick-off points at different temperatures of the piping that carry the helium gas from the tank, which is routed around the instruments for this purpose, and to the vapour-cooled shields for eventual release into free space. 

The SVM houses `warm' payload electronics on four of its eight panels and provides the necessary `infrastructure' for the satellite  such as power, attitude and orbit control, the on-board data handling and command execution, communications, and safety monitoring. It also provides a thermally controlled environment, which is critical for some of the instrument units. Finally, the SVM also provides mechanical support for the PLM, the sunshield/sunshade, a thermal shield to thermally decouple the PLM from the SVM, and it ensures the main mechanical load path during the launch.

\subsection{Telescope}
\label{sec:tel}
The {\it Herschel} telescope (Doyle et al. \cite{doyle2009}) was constructed to be as large as possible and still be compatible with no inflight deployable structures, as cold as possible with passive cooling, and providing a total wavefront error (WFE) of less than $6\,\mu$m in the focal surface interfacing to the instruments. At the same time it needed to have a low mass and the required mechanical and thermal properties. 

The optical design is that of a classical Cassegrain telescope with a 3.5\,m physical diameter primary and an `undersized' secondary, yielding an effective primary diameter of 3.28\,m. The telescope (Fig.\,\ref{fig:sc}) is constructed almost entirely ($\sim$90\% by mass) of silicon carbide (SiC). The primary mirror has been made out of 12 segments, `brazed' together to form a monolithic mirror blank, which was machined and polished. The secondary is a single piece, machined with an integral `scattercone' to suppress standing waves and the narcissus effect. The M1 and M2 optical surfaces have been coated with a high reflectivity/low emissivity  (Fischer et al. \cite{fischer2004}) aluminium layer, covered by a thin protective `plasil' (silicon oxide) coating which allows cleaning.

The telescope internal alignment and WFE performance were measured in cold conditions. Similar to ISO, a fully passive design was adopted, with no means of inflight adjustments such as e.g. focusing.

\subsection{Science instruments}
The science payload consists of three instruments, provided by consortia of institutes led by their Principal Investigators (PIs):
\begin{itemize}
\item The Photodetector Array Camera and Spectrometer (PACS, Poglitsch et al. \cite{poglitsch2010}),  PI: A. Poglitsch, Max-Planck-Institut f\"ur extraterrestrische Physik (MPE), Garching.
\item The Spectral and Photometric Imaging REceiver (SPIRE, Griffin et al. \cite{griffin2010}), PI: M. J. Griffin, Cardiff University.
\item The Heterodyne Instrument for the Far Infrared (HIFI, de Graauw et al. \cite{degraauw2010}), PI: T. de Graauw, in late 2008 succeeded by F. Helmich, SRON Netherlands Institute for Space Research, Groningen. 
\end{itemize}

\begin{table}[h]
\caption[]{Science instrument main characteristics. Acronyms relating to the detectors: superconductor-insulator-superconductor (SIS), hot electron bolometer (HEB), gallium-doped germanium (Ge:Ga), and neutron transmutation doped (NTD).}
\label{tab:instr} 
\centering
	 \begin{tabular}{ l  r }
	 \hline
	HIFI &  heterodyne spectrometer  \\
	\hline
	Wavelength coverage & 157$-$212 \& 240$-$625\,$\mu$m \\
	Field-of-view (FOV) & single pixel on sky \\
	Detectors & 5x2 SIS \& 2x2 HEB mixers \\
	Spectrometers & auto-correlator \& acousto-optical \\
	Spectral resolving power &  typically 10$^6$ \\
	\hline 
	PACS & 2-band imaging photometer  \\
	\hline
	Wavelength coverage & 60$-$85 or 85$-$130, 130$-$210\,$\mu$m \\
	Field-of-view (FOV) & 0.5$F\lambda$ sampled 1.75\arcmin x3.5\arcmin \\
	Detectors & 64x32 \& 32x16 pixel bol. arrays  \\
	\hline
	PACS &  integral field spectrometer \\
	\hline
	Wavelength coverage & 55-210\,$\mu$m \\
	Field-of-view (FOV) & (5x5 pixel) ~47\arcsec x47\arcsec \\
	Detectors & two 25x16 pixel Ge:Ga arrays \\
	Spectral resolving power  &  1000$-$4000  \\
	\hline
	SPIRE & 3-band imaging photometer  \\
	\hline
	Wavelength bands ($\lambda$/$\Delta\lambda\sim$3) & 250, 350, 500\,$\mu$m \\
	Field-of-view (FOV) & 2$F\lambda$ sampled 4\arcmin x8\arcmin \\
	Detectors & 139, 88 \& 43 pixel NTD bol. arrays  \\
	\hline
	SPIRE &  imaging fourier transf. spectrometer  \\
	\hline
	Wavelength coverage & 194$-$324 \& 316$-$671\,$\mu$m \\
	Field-of-view (FOV) & 2$F\lambda$ sampled circular 2.6\arcmin \\
	Detectors & 37 \& 19 pixel NTD bol. arrays  \\
	Spectral resolving power  &  370$-$1300 (high) / 20$-$60 (low)  \\
	\hline
	\end{tabular}
\end{table}

The three instruments complement each other (Table\,\ref{tab:instr}), enabling {\it Herschel} to offer its observers broad band photometric imaging capability in six bands with centre wavelengths of 70, 100, 160, 250, 350, and $500\,\mu$m, imaging spectroscopy over the entire {\it Herschel} wavelength coverage, and very high resolution spectroscopy over much of this range. 

A number of observing modes are provided, including point source photometry, small, and large area photometric imaging, and the observation of a single spectral line, or one or more spectral ranges, in either a single position or in various mapping modes. The actual implementations of the modes differ between the instruments, and are described in manuals provided to the observers (Sect.\,\ref{sec:gs}).

 \begin{figure*}
    \centering{
	\includegraphics[height=80mm]{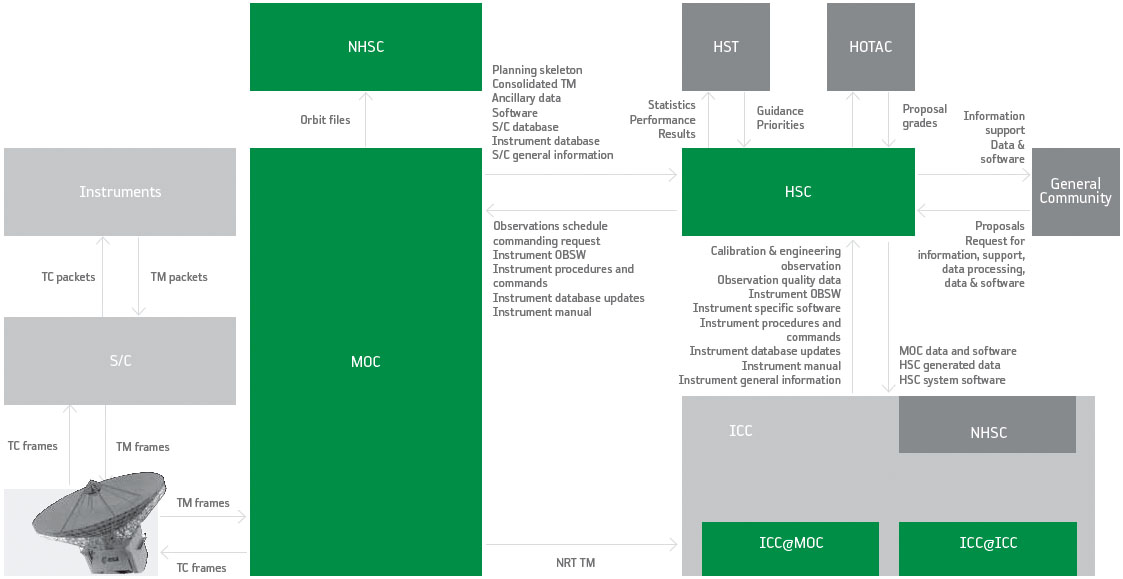}
	  }
    \caption{\label{fig:gs}
The {\it Herschel} ground segment elements (green boxes) and information flow. HST stands for the {\it Herschel} Science Team, HOTAC for the {\it Herschel} Observing Time Allocation Committee, TC for telecommand, TM for telemetry.  ICCs receive instrument TM from the HSC in their home locations (ICC@ICC), but can also `intercept' TM on location in the MOC (ICC@MOC) for near real-time (NRT) activities.
   }
\end{figure*}

\subsection{Ground segment: mission and science operations}
\label{sec:gs}
The {\it Herschel} ground segment (Fig. \ref{fig:gs}) comprises six elements:

\begin{itemize}
\item
The Mission Operations Centre (MOC), based at the European Space Operations Centre (ESOC), Darmstadt;
\item	
The {\it Herschel} Science Centre (HSC), based at the European Space Astronomy Centre (ESAC), Madrid; 
\item
Three dedicated Instrument Control Centres (ICCs), based at SRON, Groningen (HIFI), at MPE, Garching (PACS), and at Rutherford Appleton Laboratory (RAL), Didcot (SPIRE); 
\item
The NASA {\it Herschel} Science Center (NHSC), based at the Infrared Processing and Analysis Center (IPAC), Pasadena.
\end{itemize}

The MOC conducts the mission operations, it performs all contact with the spacecraft, monitors all spacecraft systems, safeguards health and safety, and performs orbit maintenance. It generates skeleton schedules for the Science Ground Segment (SGS) to `fill in' with science, calibration, and engineering observations. 
It creates and uplinks mission timelines and receives, consolidates, and provides telemetry to the SGS. 

Science operations are conducted by the SGS, a partnership among the HSC, the three ICCs, and the NHSC. 
The HSC is the interface between {\it Herschel} and (prospective) investigators in the science community. It provides information, handles calls for observing proposals, performs scientific mission planning, systematic pipeline data processing, data quality control, and populates the science archive, and offers user support related to all aspects of observing, including observation planning and proposing, observation execution, data processing and access. 

The ICCs are responsible for the successful operation of their respective instruments,  developing and maintaining instrument observing modes, and for providing specialised software and procedures for the processing of the data generated. 

The NHSC provides additional support and offers science exploitation funding for investigators based in the USA.

The SGS and the astronomical community interact through a common repository, the jointly developed scientific ground segment
software, known as the {\it Herschel} Common Science System (HCSS, Riedinger \cite{jrr2009}). The HCSS handles submission of observing proposals, scheduling of observations for execution, generation of commanding requests for the instruments and spacecraft, processing of the resulting scientific data, and the population of the archive with the generated products.

Observers are provided with the HSpot tool for observation planning, the HIPE interactive data processing software (Ott \cite{ott2010}), observing data and standard products through the {\it Herschel} Science Archive (HSA), and associated documentation including Observers' Manuals and User Guides as well as the  Helpdesk interface through the HSC website\footnote{HSC website URL: http://herschel.esac.esa.int/}.

\subsection{Integration and verification}
\label{sec:aiv}

The science instrument integration followed by the mating of the PLM and SVM was performed in 2007, before shipment to the European Space Research and Technology Centre (ESTEC) where the final spacecraft assembly and verification took place in 2008. The telescope was mated in April 2008, and in May 2008 the spacecraft level environmental verification activities commenced, including electromagnetic compatibility (EMC) tests, mechanical testing of science instrument mechanisms, such as PACS chopper and grating and SPIRE spectrometer moving mirror (requiring special spacecraft attitude), followed by acoustic and shaker testing.

Careful alignment measurements together with measured cooldown effects at telescope and payload module level were used to verify correct as-predicted optical alignment between the telescope and the instrument focal plane units inflight. In September 2008 the first System Operational Validation Test (SOVT-1) was performed, with the SGS and the MOC operating the spacecraft in a complete end-to-end test simulating four days of routine science phase  operations. The whole system was exercised from the generation of observation requests and schedules and the uplinking of mission timelines through actual execution by the spacecraft and downlinking of the generated telemetry, followed by ingestion of the data by the HSC and performing pipeline data processing and product archiving and retrieval, data propagation to the ICCs, and responding to Helpdesk questions and Targets of Opportunity.
 
After installation in the Large Space Simulator (LSS) an extensive testing campaign pertaining to both the spacecraft itself and the scientific payload took place in November-December 2008 in simulated near-inflight conditions.  A major aspect of the LSS campaign was the SOVT-2, this time extending simulations to the early mission phases before routine operations. The SOVTs together with subsequent longer duration simulations without the spacecraft established operational readiness.

\section{Observing opportunities}
\label{sec:obs} 

{\it Herschel} is an observatory; its observing time is shared between guaranteed and open time (GT and OT). The basic rules were defined in the Science Management Plan (SMP) as part of the AO for the science payload in 1997 (Sect.\ref{sec:intro}). 

In the nominal mission $\sim$20,000 hours are available for science, 32\% is GT (mainly owned by the PI consortia). The remainder is OT, which is allocated to the general community (including the GT holders) on the basis of AOs for observing time. In each AO cycle the GT is allocated first, followed by the OT.

A small amount of the open time can be allocated as discretionary time. All proposals are assessed by the {\it Herschel} Observing Time Allocation Committee (HOTAC), and all observing data are archived and will be available to the entire community after the proprietary time has passed.

\subsection{Key Programmes}
\label{sec:kps} 

Given that {\it Herschel} would not have the benefit of an all sky survey for much of its wavelength coverage, the SMP required that `Key Programmes' (KPs) in the form of `large spatial and spectral surveys' should be selected and executed early in the mission so that the results could be followed up by {\it Herschel} itself. 

\begin{figure}[!htb]
\centering{
	\includegraphics[height=5.9cm]{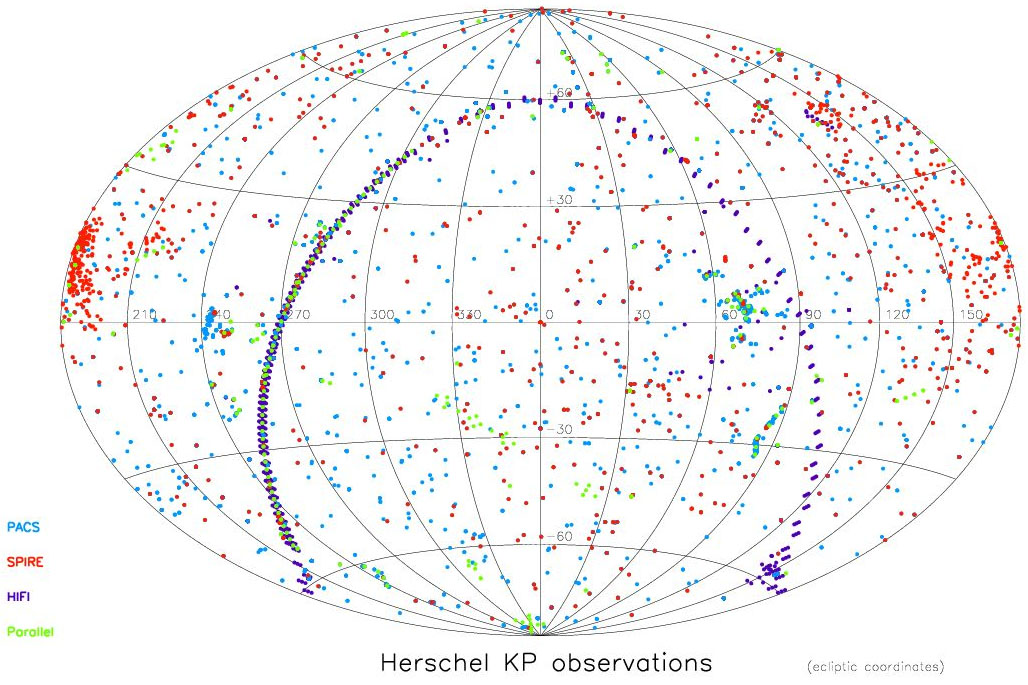}
  	}
 \caption{
 \label{fig:kps}
KP observations plotted in ecliptic coordinates, with colour-coding denoting the instrument. `Parallel' indicates performing 5-band photometric imaging using SPIRE and PACS simultaneously. 
 }
\end{figure}

An initial AO limited to KP observing proposals should therefore be issued. This process took place from February 2007 to February 2008. There were 21 GT and 62 OT proposals, out of which by coincidence also 21 were awarded observing time (Table\,\ref{tab:kps} and Fig.\,\ref{fig:kps}). 

\begin{table}[!htb]
\caption{
\label{tab:kps}
Approved KP proposals by science area. The table shows the number of proposals and the amount of time as awarded, per science category and in KP GT, KP OT, and KP in total.
} 
{\small
\begin{center}       
\begin{tabular}{|l|c|c|c|c|c|c|} \hline
Topic     & \multicolumn{2}{c|}{KP GT} & \multicolumn{2}{c|}{KP OT} & \multicolumn{2}{c|}{KP total} \\ \cline{2-7}
~~~~~~~~props             & \# & ~~hours~~ & \# & ~~hours~~ & \# & ~~hours~~ \\ \hline
Solar system         & ~1  & ~293.7  & ~1  & ~372.7  & ~2  & ~~666.4 \\ \hline
ISM/SF   & 10  & 2337.5  & 10  & 2113.2  & 20  & ~4450.7 \\ \hline
Stars                & ~2  & ~544.6  & ~0  &    0    & ~2  & ~~544.6 \\ \hline  
Gal/AGNs        & ~5  & ~983.7  & ~8  & 1930.3  & 13  & ~2914.0 \\ \hline
Cosmology            & ~3  & 1719.4  & ~2  & ~962.6  & ~5  & ~2682.0 \\ \hline
Total                & 21  & 5878.9  & 21  & 5378.8  & 42  & 11257.7 \\ \hline
\end{tabular}
\end{center}
}
\end{table} 

\subsection{Inflight AOs}
\label{sec:aos} 

The total amount of observing time allocated to KPs constitutes $\sim$55\% of the nominally available observing time. The remainder will be allocated in two additional AO cycles. The first of these is underway since February 2010,  the GT1 deadline was on 31 March 2010, while OT1 will be released on 20 May 2010, with proposal submission deadline on 22 July 2010. The final AO (GT2 \& OT2) will take place approximately a year later.

\section{Mission phases}

After the launch and early operations phase (LEOP) the inflight mission phases are:

\begin{itemize}
\item Commissioning Phase (CoP): Complete check-out of spacecraft functions and performance. Switch-on and functional verification of the instruments. Cryo-cover opening and thermal stabilisation. Nominal duration two months. 
\item Performance Verification Phase (PVP): Instrument performance, calibration, focal plane geometry, and pointing determination. Test, optimise, verify, and release the various instrument/observing modes for use. Nominal duration three months.
\item Science Demonstration Phase (SDP): Use of the released observing modes to execute selected observations from the approved KPs. Assess, optimise, and release the KPs observations for execution. Nominal duration six weeks.
\item Routine Science Phase (RSP): Execute released observations employing released observing modes. Perform engineering and routine calibration observations to optimise the observing modes and the quality of the data obtained.
\end{itemize}

The transition from CoP to PVP includes a formal review and handover of top-level responsibility to the Mission Manager. By the very nature of the activities conducted, the transitions from PVP to SDP, and from SDP to RSP are staggered by observing mode and observing programme, respectively. The RSP will continue until helium exhaustion, and will be followed by archive phases.

\section{Inflight status and performance}
\label{sec:inflight}

{\it Herschel} was launched on 14 May 2009 from Kourou at 10:12 local time (13:12 UT) at the very beginning of the launch window, by an Ariane 5 ECA launcher shared with {\it Planck}. {\it Herschel} separated from the launcher 26 min after liftoff, by then the cryogenic system had already passed a critical milestone by demonstrating nominal phase separator performance. 

\begin{figure}[!t]
\centering{
	\includegraphics[height=4.35cm]{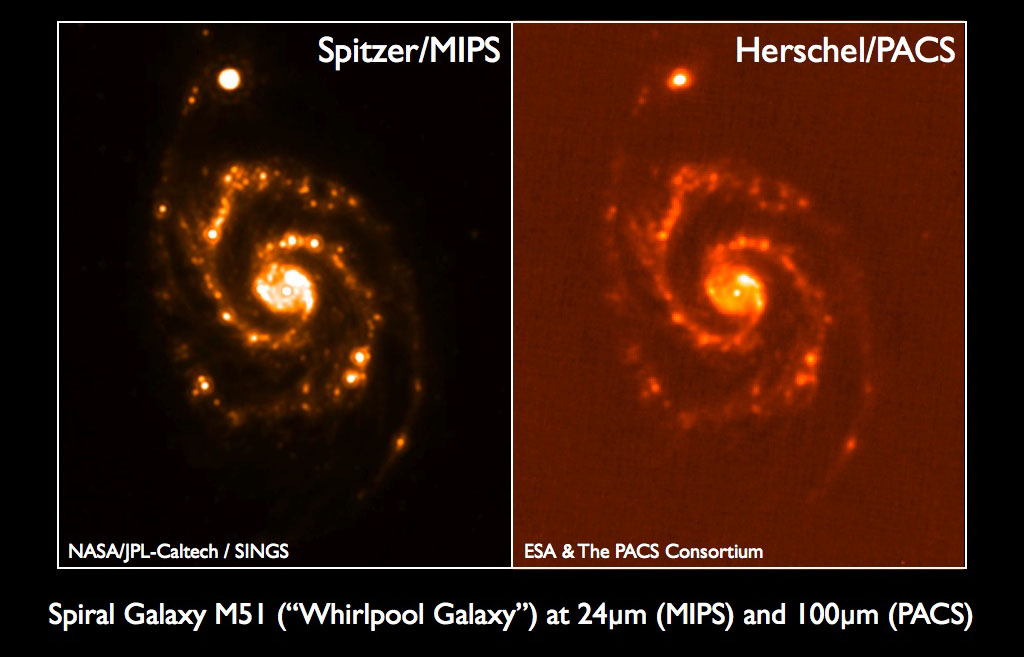}
	\includegraphics[height=4.1cm]{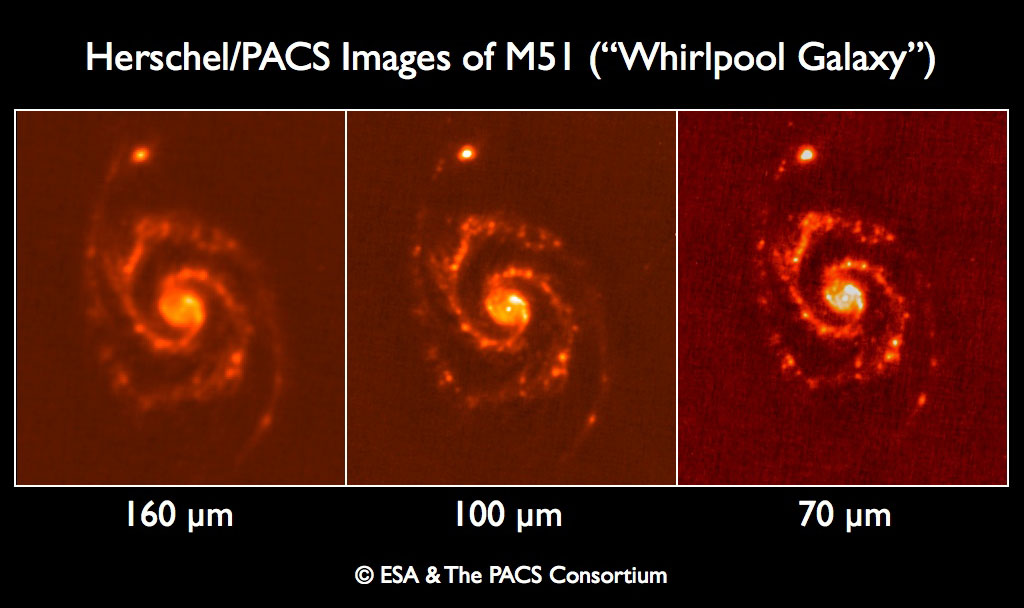}
  	}
 \caption{
 \label{fig:M51}
Top: {\it Spitzer}/MIPS and {\it Herschel}/PACS M51 images at 24$\,\mu$m and 100$\,\mu$m, respectively. Because aperture sizes and wavelengths have similar ratios and {\it Spitzer} has perfect optical performance, it can be concluded that {\it Herschel} has good optical performance at $100\,\mu$m. Bottom: {\it Herschel}/PACS images at 160, 100, and $70\,\mu$m, demonstrating the improvement of the angular resolution with shorter wavelengths. 
 }
\end{figure}

\begin{figure}[!t]
\centering{
	\includegraphics[height=6.7cm]{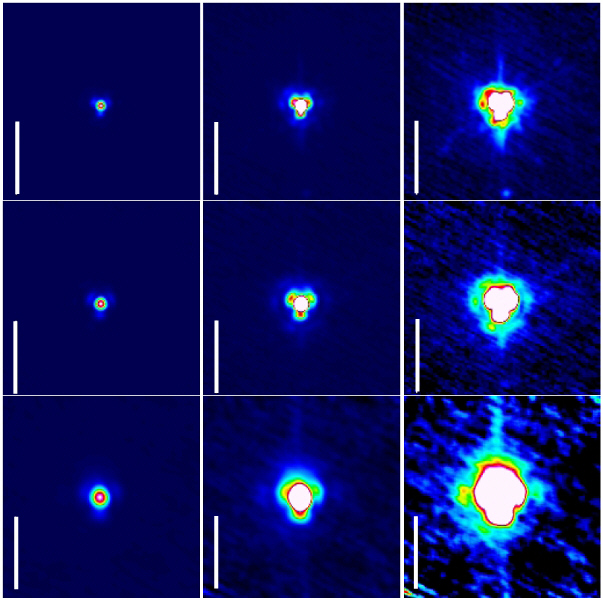}
  	}
 \caption{
 \label{fig:psf}
PSFs obtained with the PACS photometer based on observations of Vesta, obtained in scan map mode at 10\arcsec/s scan speed. The 70, 100, and 160 $\mu$m bands are arranged from top to bottom and normalised to peak, 10\%, and 1\%, from left to right. The scale bar indicates 60". Courtesy N. Geis and the PACS consortium. 
 }
\end{figure}

Commissioning activities commenced a few days into the flight. As the spacecraft cooled down, heaters kept the telescope optics at 170 K for some time (to prevent it acting as a cold trap for water vapour outgassing from the spacecraft), then allowed the
telescope to cool down in preparation for the opening of the cryo-cover. One month after the launch on 14 June 2009 the cryo-cover was manually commanded open from the MOC. Immediately after the sky became accessible for the first time test observations were carried out using the PACS photometer which were  designed to determine the relative alignment between the field-of-view (FOV) on the sky and commanded pointing, the optical load on the detectors, and to validate end-to-end the performing of observations including data processing and the generation of data products. The success was beyond expectations, and very importantly it demonstrated (Fig.\,\ref{fig:M51}) good optical performance of {\it Herschel}. The analysis of PVP observations has established the optical performance in more detail  (Fig.\,\ref{fig:psf}).

Early in PVP on 2 August 2009 the HIFI instrument went into an undefined mode, later concluded to be due to a failure in one of its warm electronics boxes, caused by radiation. With HIFI unavailable, the PVP activities had to be replanned, making more use of PACS and SPIRE. Photometric imaging by scanning the spacecraft - `scan mapping'  - turned out to produce spectacular results, and was the first observing mode to be released for use. This led to the first SDP observations being carried out in September 2009, and the first RSP observations in October 2009.  

Pointing observations of large numbers of targets have established absolute pointing performance in line with the pre-launch predictions of $\sim$2\arcsec, with a small restriction in the routinely usable solar aspect angle (SAA) from $120\degr$ to 110$\degr$ wrt telescope boresight. On 25 November 2009 the first direct measurement of the amount of remaining helium was carried out. The result, together with thermal modelling, indicates a predicted mission lifetime somewhere in the range of 3.5 to slightly above 4 years. 

After a cooperative failure investigation conducted by SRON and ESA, a plan for switching HIFI on again using redundant warm electronics units was established. HIFI was successfully reactivated on 10 January 2010.  HIFI has been allocated half the available observing time in early 2010 to enable it to be fully characterised, and its KP observations (as well as those of PACS and SPIRE)  to be consolidated, for the upcoming AO.

In summary {\it Herschel} has essentially concluded the early mission phase activities. The optical, instrument, pointing, and lifetime performances are according to expectations, and the operations concept has been validated inflight. 

In the month of the anniversary of the launch, May 2010, the {\it Herschel} First Results Symposium will feature $\sim$200 presentations of scientific results, $\sim$150 scientific papers will be in final acceptance stage, and the next open time AO will be released. In an optimistic scenario there could be three more years of {\it Herschel} observing still to come!

\begin{acknowledgements}
{\it Herschel} has only been made possible by the collective dedicated effort by a very large number of individuals over many years. This paper has been written on behalf of all those who have contributed one way or another to make {\it Herschel} the reality it now has become (see http://herschel.esac.esa.int/HerschelPeople.shtml), and for the benefit of the even more who will reap the rewards. 

The {\it Herschel} spacecraft was designed, built, tested, and launched under a contract to ESA managed by the {\it Herschel/Planck} Project team by an industrial consortium under the overall responsibility of the prime contractor Thales Alenia Space (Cannes), and including Astrium (Friedrichshafen) responsible for the payload module and for system testing at spacecraft level, Thales Alenia Space (Turin) responsible for the service module, and Astrium (Toulouse) responsible for the telescope,  with in excess of a hundred subcontractors.

HCSS, HSpot, and HIPE are joint developments
by the {\it Herschel} Science Ground Segment Consortium, consisting of ESA, the
NASA {\it Herschel} Science Center, and the HIFI, PACS and SPIRE consortia.

HIFI has been designed and built by a consortium of institutes and university departments from across Europe, Canada and the United States under the leadership of SRON Netherlands Institute for Space Research, Groningen, The Netherlands and with major contributions from Germany, France and the US. Consortium members are: Canada: CSA, U.Waterloo; France: CESR, LAB, LERMA,  IRAM; Germany: KOSMA, MPIfR, MPS; Ireland, NUI Maynooth; Italy: ASI, IFSI-INAF, Osservatorio Astrofisico di Arcetri-INAF; Netherlands: SRON, TUD; Poland: CAMK, CBK; Spain: Observatorio Astron\'omico Nacional (IGN), Centro de Astrobiolog'a (CSIC-INTA). Sweden:  Chalmers University of Technology - MC2, RSS \& GARD; Onsala Space Observatory; Swedish National Space Board, Stockholm University - Stockholm Observatory; Switzerland: ETH Zurich, FHNW; USA: Caltech, JPL, NHSC.

PACS has been developed by a consortium of institutes led by MPE
(Germany) and including UVIE (Austria); KU Leuven, CSL, IMEC (Belgium); CEA,
LAM (France); MPIA (Germany); INAF-IFSI/OAA/OAPD, LENS, SISSA
(Italy); IAC (Spain). This development has been supported by the funding
agencies BMVIT (Austria), ESA-PRODEX (Belgium), CEA/CNES (France),
DLR (Germany), ASI/INAF (Italy), and CICYT/MCYT (Spain).

SPIRE has been developed by a consortium of institutes led by
Cardiff University (UK) and including Univ. Lethbridge (Canada);
NAOC (China); CEA, LAM (France); IFSI, Univ. Padua (Italy); IAC
(Spain); Stockholm Observatory (Sweden); Imperial College London,
RAL, UCL-MSSL, UKATC, Univ. Sussex (UK); and Caltech, JPL, NHSC,
Univ. Colorado (USA). This development has been supported by
national funding agencies: CSA (Canada); NAOC (China); CEA,
CNES, CNRS (France); ASI (Italy); MCINN (Spain); SNSB
(Sweden); STFC (UK); and NASA (USA). 

\end{acknowledgements}



\begin{thebibliography}{}
      
\bibitem[2009]{doyle2009} Doyle, D., Pilbratt, G., Tauber, J. 2009,
      Proc IEEE 79, 1403
      
\bibitem[2004]{fischer2004} Fischer, J., Klaassen, T., Hovenier, N. et al. 2004,
     Appl. Opt. 43, 3765

\bibitem[2003]{frisk2003} Frisk, U., Hagstr\"om, M., Ala-Laurinaho, J. et al. 2003,
      A\&A, 402, L27
      
\bibitem[2010]{degraauw2010} de Graauw, T., Helmich, F.P., Phillips, T.G. et al. 2010,
      A\&A, this volume
      
\bibitem[2010]{griffin2010} Griffin, M.J., Abergel, A., Abreu, A. et al. 2010,
      A\&A, this volume
     
\bibitem[1800]{herschel1800} Herschel, W. 1800,
Phil. Trans. Roy. Soc. London 90, 255

\bibitem[1996]{kessler1996} Kessler, M.F., Steinz, J.A., Anderegg, M.E. et al. 1996,
      A\&A, 315, L27
  
\bibitem[1984]{longdon1984} Longdon, N. (ed) 1984, European Space Science: Horizon 2000, ESA SP-1070
      
\bibitem[2000]{melnick2000} Melnick, G.J., Stauffer, J.R., Ashby, L.N. et al. 2000, 
      ApJ, 539, L77 
      
\bibitem[2007]{murakami2007} Murakami, H., Baba, H., Barthel, P. et al. 2007,
      PASJ, 59, S369
      
\bibitem[1984]{neugebauer1984} Neugebauer, G., Habing, H. J., van Duinen, R. et al. 1984, 
       ApJ, 278, L1
       
\bibitem[2003]{nordh2003} Nordh, H.L., von Sch\'eele, F., Frisk, U. et al. 2003,
      A\&A, 402, L21
      
\bibitem[2010]{ott2010} Ott, S. 2010, in ASP Conference Series, Astronomical Data Analysis
Software and Systems XIX, Y. Mizumoto, K.-I. Morita, and M. Ohishi, eds., in press
      
\bibitem[2001]{glp2001} Pilbratt G.L., Cernicharo J., Heras, A.M. Prusti T. \& Harris R. (eds.) 2001, The Promise of the {\it Herschel} Space Observatory, ESA SP-460
      
\bibitem[2010]{poglitsch2010} Poglitsch, A., Waelkens, C., Geis, N. et al. 2010,
      A\&A, this volume
      
\bibitem[2009]{jrr2009} Riedinger, J.R. 2009, 
	ESA Bullentin, 141, 12

\bibitem[1997]{mrr1997} Rowan-Robinson M., Pilbratt G., Wilson A. (eds.) 1997, The Far Infrared and Submillimetre UniverseÕ, 
	ESA SP-401

\bibitem[2004]{werner2004} Werner, M.W., Roellig, T.L., Low, F.J. et al. 2004,
      A\&A, 315, L27
      
\end{thebibliography}
\end{document}